# Transport evidence of triply degenerate nodal semimetal YRh$_6$Ge$_4$

Yanglin Zhu[1], Xin Gui[2], Yu Wang[1], David Graf[3], Weiwei Xie[2], Zhiqiang Mao[1*]

[1] Department of Physics, Pennsylvania State University, University Park, PA 16802

[2] Department of Chemistry, Louisiana State University, Baton Rouge, LA 70803

[3] National High Magnetic Field Laboratory, Tallahassee, FL 32310

## Abstract

We have investigated magnetotransport properties of YRh$_6$Ge$_4$, which was recently predicted to be a triply degenerate nodal semimetal. We find it exhibits remarkable signatures of a chiral anomaly, manifested by large negative longitudinal magnetoresistance, quadratic field dependence of magnetoconductance and planar Hall effect. Furthermore, we have also observed Shubnikov-de Haas (SdH) quantum oscillations in the magnetoresistivity measurements on this material. The analyses of the SdH data reveal two point-like Fermi surfaces and these pockets are found to host nearly massless fermions. The small size of these Fermi pockets is in a good agreement with the theoretical prediction that the triply degenerate point in YRh$_6$Ge$_4$ is much closer to the Fermi level than previously demonstrated triply degenerate nodal semimetals such as MoP and WC. These results suggest YRh$_6$Ge$_4$ may serve as a model system to probe exotic properties of three-component fermions and understand their underlying physics.

*email: zim1@psu.edu

## I. INTRODUCTION

Topological semimetals (TSMs) are characterized by topologically protected band crossings near the Fermi level (FL), which leads to many exotic properties such as large magnetoresistance [1], high carrier mobility [1,2], chiral anomaly [3–7], and intrinsic anomalous Hall effect [8–12]. TSMs can be categorized by the band degeneracy at crossing points. Three-dimensional (3D) Dirac semimetals (DSMs) feature four-fold degenerate band crossing nodes (i.e. Dirac nodes), which were first theoretically predicted and then experimentally observed in $Na_3Bi$ [13,14] and $Cd_3As_2$ [15–18]. When the spin degeneracy is lifted by breaking time-reversal symmetry or inversion symmetry, a DSM evolves into a Weyl semimetal (WSM), which is characterized by non-degenerate bands crossing, with each crossing point (i.e. Weyl node) having two-fold degeneracy [19,20]. WSMs were first demonstrated in TaAs-class materials [20–26]. In addition to DSMs and WSMs, other forms of TSMs with three-, six-, and eight-fold degenerate nodal points have been also proposed [27–32]. The three-degenerate nodal point TSM has been predicted in many materials such as WC- type families, including WC [31,33], ZrTe [33,34], MoP [35] and TaN [36]), and probed by ARPES in MoP [37] and WC [38]. In these materials, their band structures show band crossings between a doubly degenerate and a non-degenerate band near the FL. Such band crossings are protected by the combination of rotation and mirror symmetry [31,33,34,36,39]. Other materials predicted to have triply degenerate nodal points include $Li_3NaN$ [40], LaPtBi [41], $NaCu_3Te_2$ [42,43], ZrO [44], $APd_3$(A=Sn, Pb) [45], $TiB_2$ [46,47], $Cu_3TeO_6$ [48], GdN [49], TaS [50], $PtBi_2$ [51], MoC [52], carbon honeycombs (CHCs) [53]. All these predictions are still waiting for experimental verifications.

Materials with triply degenerate fermions are expected to exhibit properties distinct from DSMs and WSMs. For instance, they carry net Berry flux $|v| = 2$, leading to two surface Fermi arcs connecting the surface projections of triply degenerate points. When a magnetic field is applied, the Zeeman effect splits each 3-fold degenerate node into Weyl points, resulting in a topological phase transition. The chiral anomaly is also expected for triply degenerate nodal semimetals but shows different characteristics in comparison with WSMs. The negative longitudinal MR (LMR) induced by the chiral anomaly in triply degenerate nodal semimetals occurs only when the current is applied to the $C_3$ rotation axis. Among the predicted triply degenerate nodal TSMs, the chiral anomaly induced negative LMR is observed only in WC thus far [54]. Recently, intermetallic compounds $R$Rh$_6$Ge$_4$ ($R$=Y, La, Lu) have been predicted to host triply degenerate points in their band structures [55]. These materials crystallize in the hexagonal structure with space group $P\bar{6}m2$, as shown in Fig. 1a. Compared to previously demonstrated triply degenerate nodal semimetals, $R$Rh$_6$Ge$_4$ is found to have triply degenerate points much closer to the FL (within a range of 50meV from the Fermi level, contrasted to the 200 meV value in WC [38]). Therefore, $R$Rh$_6$Ge$_4$ provides an excellent platform to probe exotic properties of triple-component fermions. In this article, we report on the transport evidence of triply degenerate fermions of YRh$_6$Ge$_4$. We not only observed chiral anomaly induced negative LMR and planar Hall effect, but also probed the point-like Fermi pockets hosting triple-component fermions through Shubnikov-de Haas (SdH) quantum oscillations. Our findings establish a promising platform for exploring new exotic properties of three-component fermions and understanding their underlying physics.

## II. EXPERIMENTAL

Single crystal YRh$_6$Ge$_4$ was synthesized through the flux method [56]. The Y pieces, Rh, Ge powder and Bi granule were mixed with molar ratio 1:5:4:20 and loaded into an Al$_2$O$_3$ crucible, then sealed in a quartz tube under vacuum. The mixture was then heated up to 1050 °C and held at this temperature for 48 hours for homogeneously melting, followed by a slow cooling down to 750 °C at a rate of 2 °C per hour and then a quick cooling down (4°C/h) from 750 °C to 550 °C. Black rod-like crystals (Fig. 1b) can be obtained after removing the Bi flux by centrifugation.

To confirm the crystal structure of synthesized crystals, we performed single crystal X-ray diffraction measurements on a crystal with the dimensions of ~15× 15× 20 μm$^3$ at room temperature using a single crystal diffractometer, Bruker Apex II (Mo radiation). We found our YRh$_6$Ge$_4$ crystals indeed have a hexagonal structure with the space group of $P\bar{6}m2$. In Fig. 1c and 1d, we present the diffraction patterns of the (h0l) and (hk0) planes. All circled diffraction spots on these two scattering planes can be indexed with the hexagonal structure. The detailed analyses of these diffraction patterns yield the lattice parameters of $a$= 7.067(3)Å and $c$= 3.862(2)Å), consistent with those previously reported in literature [56]. Furthermore, we also observed satellite diffraction spots corresponding to a superlattice, i.e. those weak spots between circled spots in Fig. 1c and 1d. The twinning assumption has been well examined and we can exclude the possibility of extra reflections due to crystal twinning. These weak spots cannot be indexed with the commensurate supercell structure of the previously reported LaRh$_6$Ge$_4$-type structure [56]. The Q-vector of the supercell structure extracted from Fig. 1c and 1d is ~ 0.176, suggesting an incommensurate superlattice. Because of the presence of such superlattice reflections, the crystal structure cannot be refined based on our current measurements. The origin of such an incommensurate superlattice is yet to be clarified. We conducted systematic magnetotransport measurements on YRh$_6$Ge$_4$ single crystals using a standard four-probe method in a Physical Property Measurement System (PPMS, Quantum Design) and high-field field measurements were carried out at the National High Magnetic Field Laboratory (NHMFL) in Tallahassee.

## III. RESULTS AND DISCUSSIONS

Figure 2 shows the transport properties of YRh$_6$Ge$_4$ single crystals measured by PPMS. In these measurements, the electrical current was applied along the axial direction of the rod, which is the $c$ axis of the crystal (Fig. 1b). YRh$_6$Ge$_4$ exhibits metallic behavior in the temperature dependence of resisitivity, but its residual resistivity shows strong sample dependence. Fig. 2a presents the resistivity data at zero field of three typical samples, labelled by S1, S2 and S3. Their residual resistivity is 0.04 mΩ.cm, 0.03 mΩ.cm and 0.02 mΩ.cm respectively. These samples exhibit very different magnetotransport behavior and the large negative LMR associated with the chiral anomaly is observed only in S1-type samples. These differences can possibly be attributed to different chemical potential among these three types of samples, which will be discussed in great details below. We will first focus on discussing the properties of the S1 sample and compare them with those of the S2 and S3 samples at the end. From Fig. 2b, it can be seen that the resistivity $\rho_{xx}$ of S1 becomes weakly temperature dependent below 20K with a slight upturn under zero field. The application of magnetic field along the $c$-axis strongly suppresses $\rho_{xx}$ for $T < 20$K, indicating negative LMR. Field sweeps of magneotresistivity (defined as MR = $[\rho(B)-\rho(0)]/\rho(0)$) at various fixed temperatures are presented in Fig. 2c, from which we find the MR becomes remarkably negative below 15K (about -5% at 9T and 2K), but positive above 15K, with a cusp-like feature at zero field. The cusp-like feature can be attributed to weak-antilocalization behavior.

Given YRh$_6$Ge$_4$ is predicted to possess a triply degenerate nodal point close to the FL, the most possible origin of the observed negative LMR is the chiral anomaly. This is indeed verified through our detailed experiments as described below. Before showing other experimental results, it should be pointed out that negative LMR can also be due to a current jetting effect for a sample with high mobility [57]; however, this scenario usually occurs in a sample with a small aspect ratio. Since our sample is rod-like and the aspect ratio of the samples used in our experiments is large, ~10, the current jetting effect can be excluded. The chiral anomaly origin of our observed negative LMR is first demonstrated by the angular dependence of MR shown in Fig. 2d where we find negative MR is gradually suppressed when the field is rotated away from the current direction. MR becomes positive when the field tilt angle $\theta$ is above 12° and the weak-antilocalization behavior becomes much more significant correspondingly. Since the chiral anomaly originates from the charge pumping bewteens paired Weyl cones with opposite chirality and the resulting topological current responsible for the chiral anomaly is proportional to $\boldsymbol{E}\bullet\boldsymbol{B}$ where $\boldsymbol{E}$ and $B$ represent electric and magnetic fields respectively [7,58], our observed angular dependence of LMR in YRh$_6$Ge$_4$ is in a good agreement with such a mechanism. Furthermore, we also find the non-oscillatory component of magnetoconductivity (i.e. the inverse of $\rho_{xx}$ for $B//I$) of the S1 sample follows $\boldsymbol{B}^2$ dependence (inset, Fig. 2b), consistent with the theroetically-predicted scaling behavior of magnetoconductance stemming from the chiral anomaly [7,58]. We note a similar $\boldsymbol{B}^2$ dependence of magnetoconductance has been demonstrated in WSMs such as TaP [59] and GdPtBi [60].

In general, the chiral anomaly in WSMs can also lead to another exotic phenoemnon - planar Hall effect (PHE) [61–64], which refers to the appearance of Hall voltage when $\boldsymbol{E}$ and $\boldsymbol{B}$ are coplanar. To further corroborate the chiral anomaly in YRh$_6$Ge$_4$, we carried out PHE measurements on this material. The data obtained from these measurements are presented in the supplementary Fig. S1, from which the planar Hall resistivity $\rho_{xy}^{PHE}$ is found to show a 2-fold symmetry with the in-plane rotation of magnetic field. However, we observed a clear deviation from the sin($2\varphi$) dependence expected for the PHE of WSMs, which can be attributed to the involment of the $\rho_{xx}$ component caused by the asymmetry of Hall contacts, which cannot seperated from $\rho_{xy}$.

As noted above, for triply degenerate nodal semimetals, a chiral anomaly is present only when both the current and magnetic field are applied to the $C_3$-rotation axis and this has been demonstrated in WC [54]. For YRh$_6$Ge$_4$, since its $C_3$-rotation axis is along the *c*-axis (Fig. 1a and 1b), our experimetal set-up for LMR measurements (Fig. 2c and 2d) satisfies the conditions for observing chiral anomaly, so it is not surprising to oberve the negative LMR in our experiments. However, the rod-like crystal does not allow us to apply current along other crystographic directions so that we could not check if the chiral anomaly is absent when the current and magnetic field are not along the c$_3$-rotation axis. In addition to negative LMR, we also observed clear SdH oscillations. The systematic analyses of SdH oscillations will be given in a later section.

To further explore the exotic quantum transport properties of YRh$_6$Ge$_4$, we performed high-field magnetotransport measurements in the NHMFL. Fig. 3a displays the high-field LMR data under various field orientation angles $\theta$, which were taken using a 31T magnet. The variation of LMR with $\theta$ is consistent with the data taken in the PPMS (Fig. 2d). Importantly, from these data, we found that the negative LMR continues to grow until the field is increased to 20T, reaching ~ -14% near 20 T. Above 20T, the SdH oscillations probed in the low field range vanish and the LMR exhibits a plateau-like feature. This feature was made much clear in the measurements conducted in the 45T hybrid magnet which allows measurements in the 11-45T field range. In Fig. 3b, we put together the data taken in the 31T and 45T magnets for a few field orientation angles. These data clearly show the plateau for *B//I* ($\theta = 0°$) extends to ~35T, beyond which LMR displays a steep drop. The tilt of magnetic field has a strong effect on the LMR drop near 35T. When $\theta$ is increased to 7°, the drop near 35T almost disappears, but the plateau extends to a much greater field range (20-40T). This plateau as well as the drop near 35T may refelct new exotic phenomena in the quantum limit, or originate from SdH oscillations of another larger Fermi pocket, as will be discussed below.

The observation of the chiral anomaly in YRh$_6$Ge$_4$ suggests a possible Weyl state emerging under magnetic field. As indicated above, theory predicts that triply-degenerated nodes could split into Weyl nodes by the Zeeman effect when the magnetic field is applied along the C$_3$ symmetry axis [38,54]. All signatures related to the chiral anomaly seen in our experiment agrees well with this theoretical scenario. As indicated above, among all the previously-predicted triply degenerate nodal TSMs, WC is the only material which was found to show the chiral anomaly induced negative LMR. This material hosts multiple triply degenerate nodes; the one which is the nearest to the FL is located at ~200 meV below $E_F$. In contrast, the triply degenerate nodes in YRh$_6$Ge$_4$ is much closer to $E_F$ according to the band structure calculations, ~50 meV above $E_F$ [55]. Our analyses of SdH oscillations provide strong support for this prediction, as will be discussed below.

As seen in Fig. 2c, the SdH oscillations in YRh$_6$Ge$_4$ start to emerge from ~1T; it decays very fast when the magnetic field is rotated from parallel to perpendicular to the current direction and disappears when the field tilt angle $\theta$ is increased above 17° (Fig.2d), indicating highly anisotropic energy bands. The Fast Fourier transform (FFT) analyses of the oscillation pattern with the background being subtracted (Fig. 4a) reveal two oscillation frequencies, i.e. $F_\alpha$ = 2T and $F_\beta$ = 6.8T, as shown in Fig. 4b. From the fits of the temperature dependences of the FFT oscillation amplitudes by the temperature damping factor of the Lifshitz-Kosevich formula, $R_T = \alpha T m^*/[m_0 B \sinh(\alpha T m^*/m_0 B)]$ where $\alpha = (2\pi^2 k_B m_0)/(\hbar e)$ (Fig. 4c), the effective mass $m^*$ is estimated to be 0.013 $m_0$ and 0.015 $m_0$ ($m_0$, free electron mass), respectively, for the $F_\alpha$- and $F_\beta$- bands, indicating the quasiparticles hosted by $F_\alpha$ and $F_\beta$ bands in YRh$_6$Ge$_4$ are nearly massless. We note that the value of $m^*$ extracted from the fit of the temperature dependence of FFT amplitude depends on the range of magnetic field used for FFT analyses in some cases [65]. The $m^*$ values given above for YRh$_6$Ge$_4$ are estimated from the analyses of the oscillation pattern in the 0.3-20T field range. We also performed the FFT analyses for the sdH oscillations in the 0.4-9T range, probed in the measurements by the PPMS. $m^*$ extracted from these analyses is 0.012 $m_0$ and 0.013 $m_0$ for the $F\alpha$- and $F\beta$-bands, respectively, comparable to the $m^*$ derived from the analyses in the 0.3-20T field range.

From the quantum oscillation frequencies extracted above, we can also evaluate the extremal cross-section area $A_F$ of the Fermi surface comprised of the $F_\alpha$ and $F_\beta$ bands using the Onsager relation $F = (\Phi_0/2\pi^2)$. The frequency of $F_\alpha$ = 2.0T and $F_\beta$ = 6.8T correspond to $A_{F,\alpha}$ = 0.019 nm$^{-2}$ and $A_{F,\beta}$ = 0.065 nm$^{-2}$ respectively. Such small values of $A_F$ indicate very small Fermi surfaces. From comparison with the calculated band structure and Fermi surfaces of YRh$_6$Ge$_4$ [55], we infer that the two calculated small electron pockets at point A at the Brillouin zone boundary (Fig. 6b in [55]) should be comprised of the $F\alpha$- and $F\beta$-bands probed in our experiments. Given the quantum oscillation frequencies of these two bands are so low, their quantum limit should be reached above 15T, which can explain the vanishing of the sdH oscillations associated with these two bands above 15T. Regarding the magnetoresistance's plateau in the high-field regime (20-35T) as well as its drop above 35T, there are two possible origins. One is that it may reflect a new quantum state emerging in the quantum limit. Theory predicts the quantum limit could possibly incur ordered states such as a charge-density wave or spin-density wave [66–68]. However, we cannot tell if such states occur to YRh$_6$Ge$_4$ in its quantum limit state only in terms of our current data. The other possibility is that the magnetoresistance's drop near 35T originates from the SdH oscillations caused by other larger Fermi pockets. Band structure calculations have shown the existence of one large electron pocket and one large hole pocket besides two small electron pockets hosting three-component fermions [55]. High-field measurements above 45T are needed to verify if this is the case, which is beyond the scope of this work.

Finally, let's compare the magnetotransport properties of sample S1 with those of samples S2 and S3. The MR data of samples S2 and S3 at 2K under various field orientation are presented in Fig. 4d and 4f respectively. Sample S2 also exhibits negative MR for $\theta <12°$ and remarkable weak-antilocalization behavior for $\theta > 12°$, but its magnitude of LMR (~ 2% even at 30T ) is much smaller than that of sample S1 (~ 13% at 30T). SdH oscillations are also observed in S2, but its oscillation pattern looks very different from that of S1 (see Fig. 4a) and its oscillation frequencies derived from the FFT analyses are $F_1$=8T and $F_2$=21T respectively, as shown in Fig. 4e where the FFT spectrum of S1 at 2K is also included for comparison. For S3, its negative LMR is very small (<1%); when the field is above 9T, its MR becomes positive. The weak-antilocalization seen in S1 and S2 also disappears in S3. Moreover, SdH oscillations also become barely observable in S3 (Fig. 4f). These observations imply that, although the band structure calculations [55] show YRh$_6$Ge$_4$ has triply degenerate points at ~50 mV above the FL, in the real synthesized crystals, the chemical potential is sample dependent and may be away from the calculated FL for some samples due to the self-doping caused by non-stoichiometric chemical composition. In fact, the stoichiometric control in bulk crystal growth has been known as a challenging problem which is hard to be overcome. The crystal growth of YRh$_6$Ge$_4$ has apparently encountered such a problem. For S1, the chemical potential is supposed to be close to the theoretical calculated $E_F$ in ref [55], since its SdH oscillations probe the two calculated small electron pockets hosting three-component fermions as discussed above. However, in S2, its SdH oscillations frequencies do not show the $F_\alpha$ =2T component, but only the $F_1$=8T and $F_2$=21T components, implying its chemical potential should be lower than that of S1 so that the $F_\alpha$ band is not occupied. The $F_1$=8T component should arise from the $\beta$ pocket, while the $F_2$=21T likely stems from the trivial electron pocket. The chemical potential of S3 should be much lower than those of S1 and S2 such that its magnetotransport properties are dominated by the trivial bands.

## IV. CONCLUSION

In summary, we have synthesized the single crystals of YRh$_6$Ge$_4$ and performed systematic magnetotransport studies on this material. We observed remakable signatures of a chiral anomaly which can be attribed to the topological phase transition from the triply degenerate nodal semimemtal state to the Weyl semimetal state. Furthermore, we also probed two point-like electron pockets through SdH oscillations, which agrees well with the two calculated small electron pockts which host three-component fermions. These results also demonstrate that the triply degenerate nodal points in YRh$_6$Ge$_4$ are indeed much closer to the FL than those in previously-established triply degenerate nodal semimetals such as MoP and WC. Therefore, our work establish a new promising playground for probing new exotic properties of triply degenerate nodal semimetal states and understanding their underlying physics.

**Acknowledgement**

This work is supported by the US National Science Foundation under grants DMR 1917579 and 1832031. This work at LSU is supported by Beckman Young Investigator (BYI) award. A portion of this work was performed at the National High Magnetic Field Laboratory, which is supported by National Science Foundation Cooperative Agreement No. DMR-1157490 and the State of Florida.

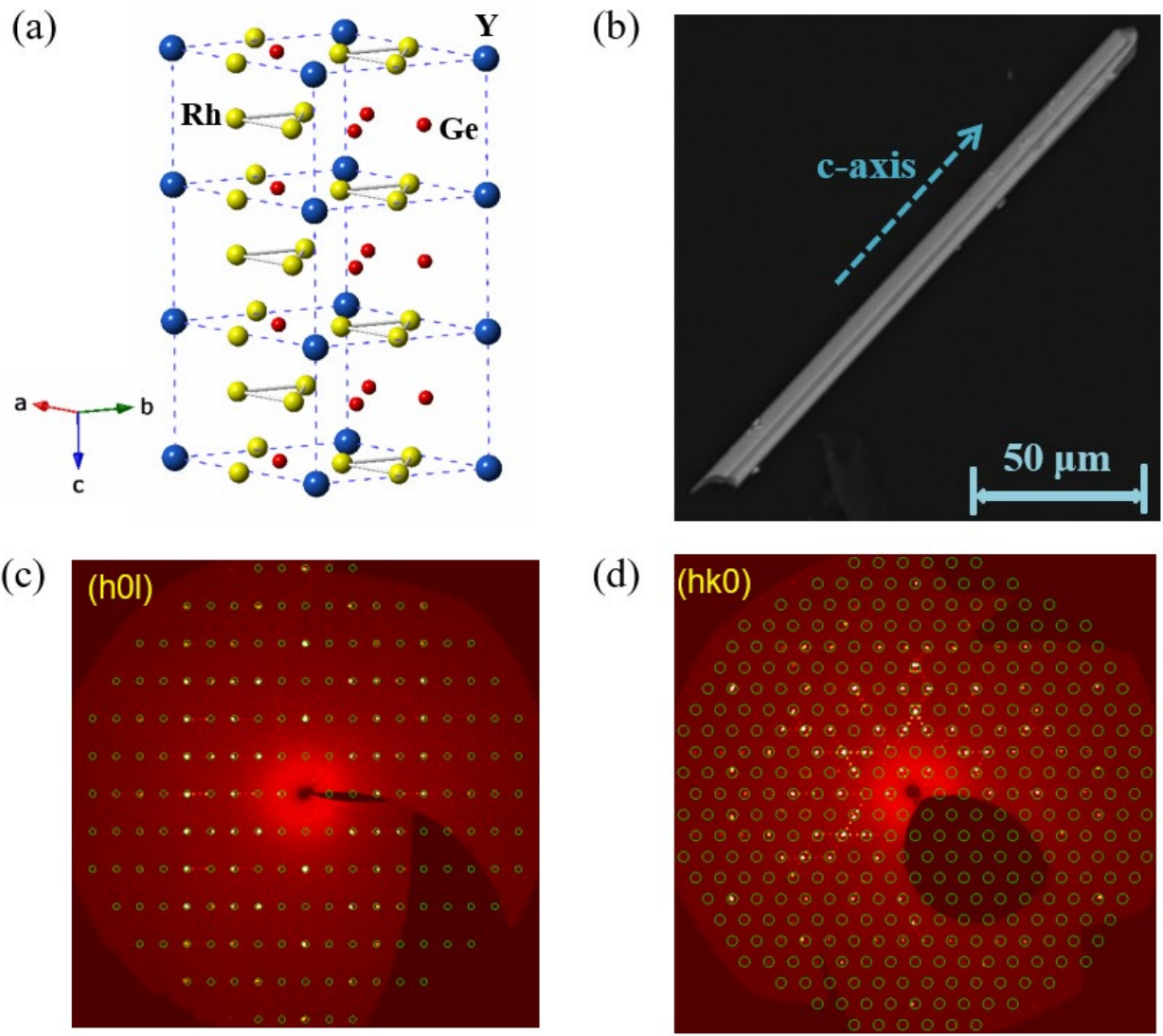

Figure 1. (a) Crystal structure of YRh$_6$Ge$_4$. (b) A crystal image of YRh$_6$Ge$_4$. (c) and (d) Single crystal X-ray diffraction precession image of the (h0l) and (hk0) planes in the reciprocal lattice of YRh$_6$Ge$_4$ at 300K. The strong intensity spots can be fitted with the LaRh$_6$Ge$_4$-type crystal structure.

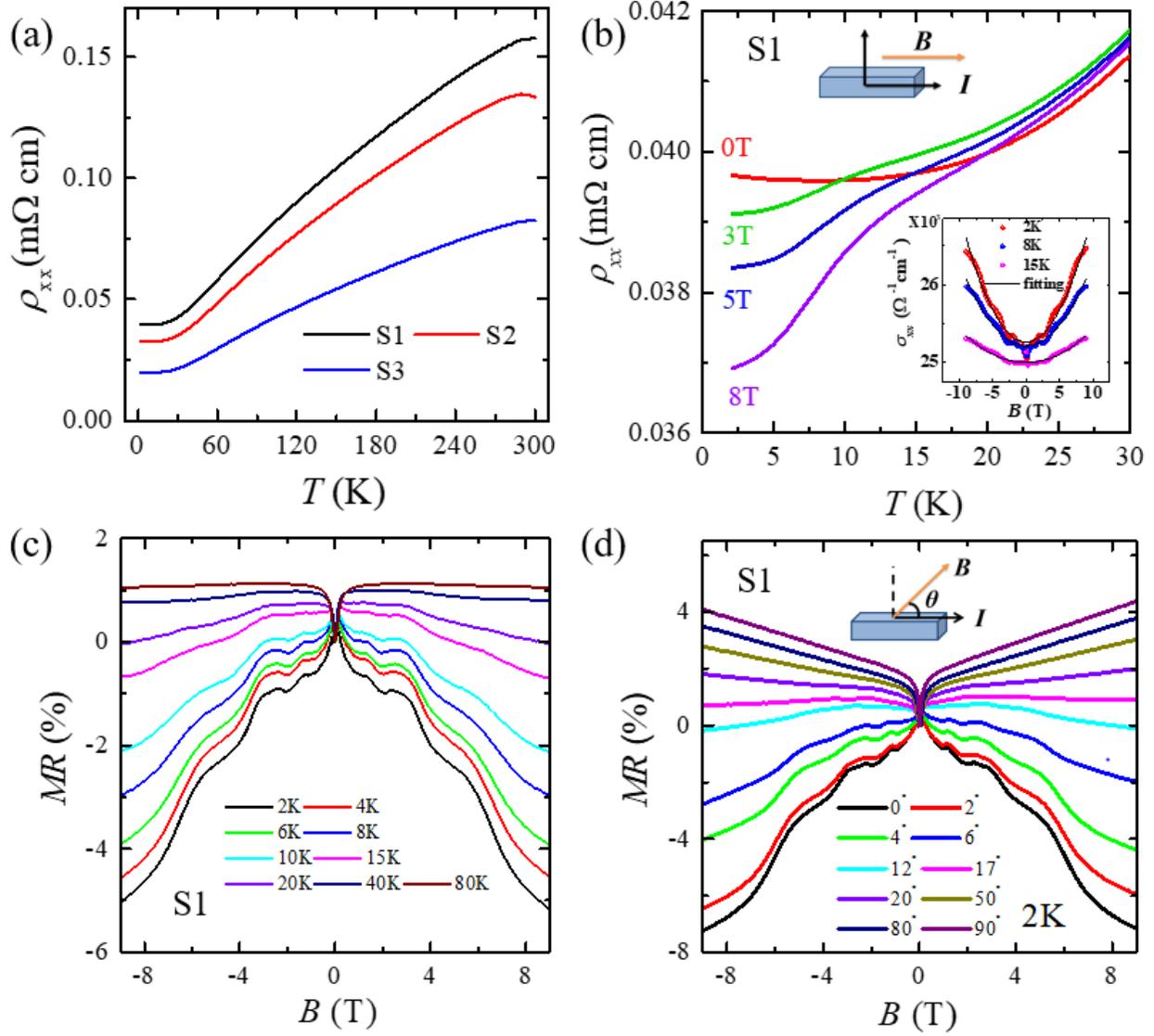

Figure 2. (a) Temperature dependence of longitudinal resistivity under zero magnetic field for three different samples. (b) Temperature dependence of longitudinal resistivity under various magnetic fields for sample S1. Inset shows the field dependence of magnetoconductivity at various temperatures for S1; the black solid lines represent the fits to the $B^2$ dependence. (c) Field dependence of longitudinal magnetoresistivity $\Delta\rho/\rho_0 = [\rho(B)-\rho(B=0)]/\rho(B=0)$ at various temperatures for sample S1. (d) Field dependence of magnetoresistivity at 2K under various field orientations measured in low 0-9T field range.

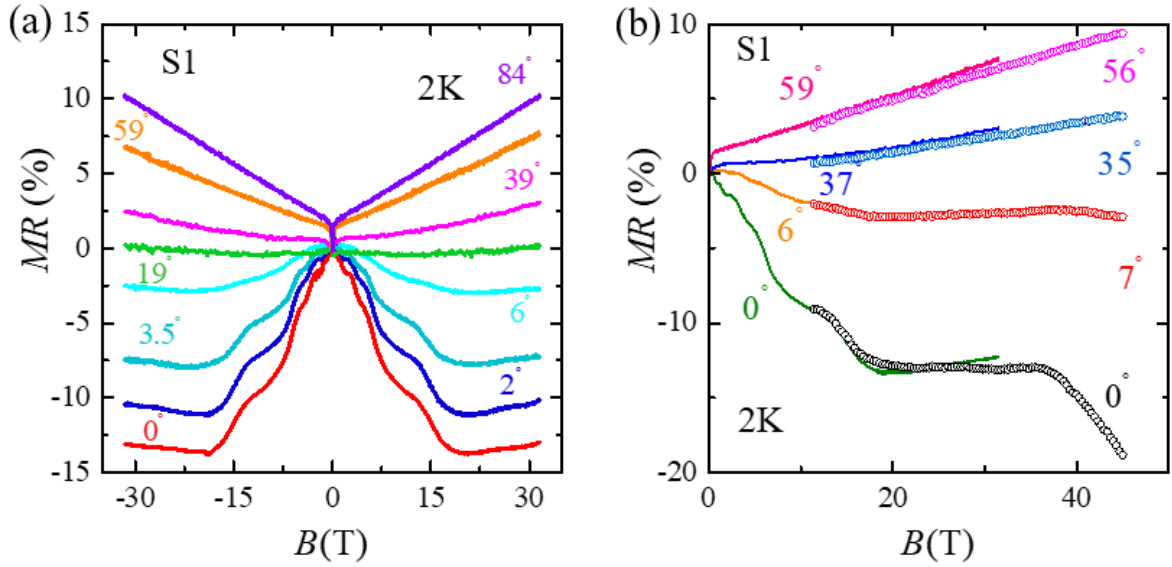

Figure 3. (a) Field dependence of magnetoresistivity at 2K under various field orientations measured using the 31T magnet system at the NHMFL. (b) Field dependence of magnetoresistivity at 2K measured under a few field orientations in both the 31T and 45 T magnet systems for S1.

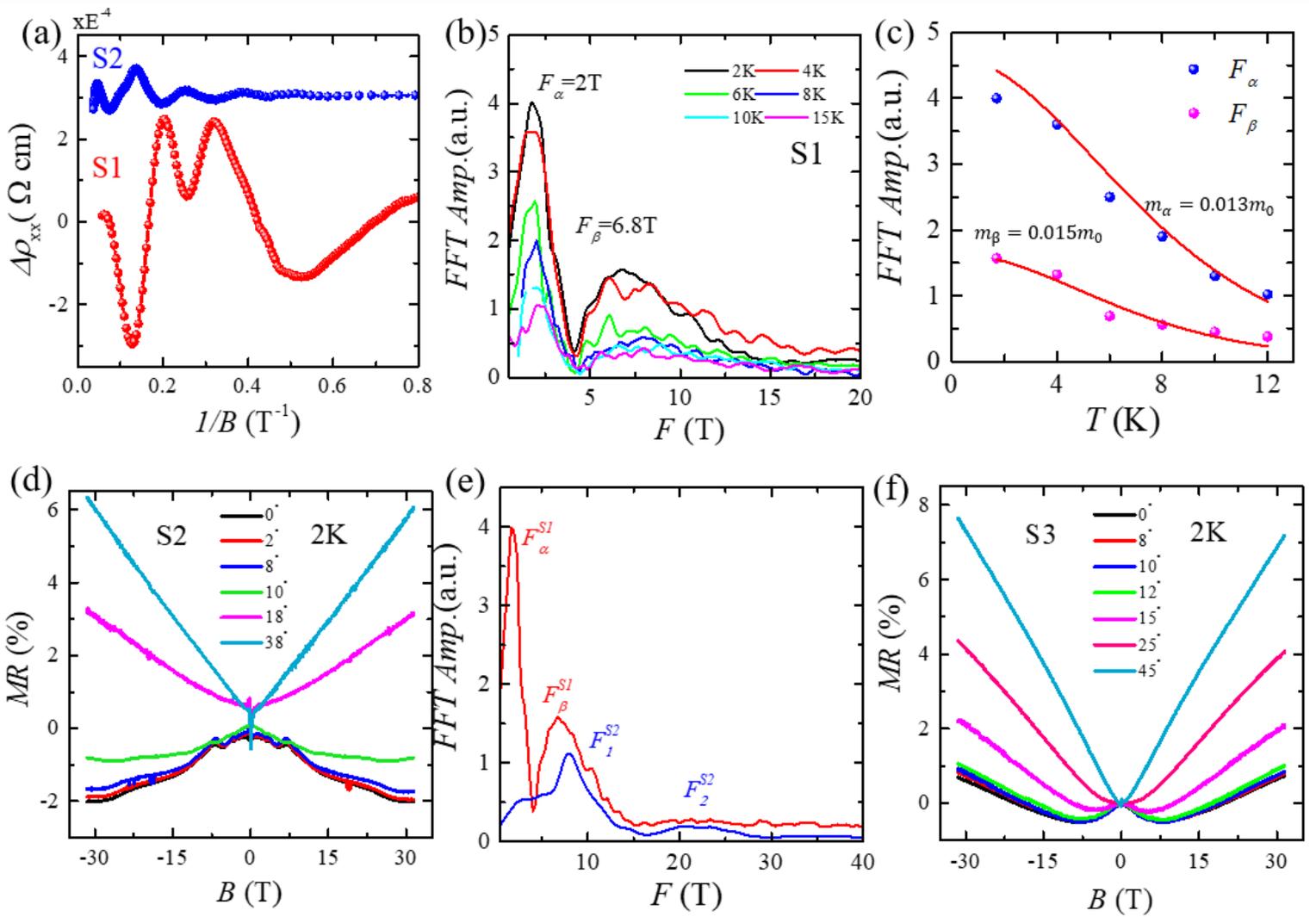

Figure 4. (a) The SdH oscillation patterns after subtracting non-oscillation background for S1(red curve) and S2 (blue curve). The data of S2 have been shifted for clarity. (b) FFT spectra of the SdH oscillations for $B // I$ for S1. (c) The fits of the FFT amplitudes of the SdH oscillations by the temperature damping factor $R_T$ in the LK formula. (d) and(f) field dependences of magnetoresistivity at 2K under various field orientations for S2(d) and S3(f). (e) The FFT spectra of the SdH oscillations at 2K for S1 and S2.